\documentclass[1p,times]{elsarticle}

\usepackage{subcaption}
\usepackage{amsmath} 
\usepackage{amssymb}
\usepackage{lipsum}

\usepackage{graphicx}
\usepackage{dcolumn}
\usepackage{bm}

\DeclareOldFontCommand{\rm}{\normalfont\rmfamily}{\mathrm}
\DeclareOldFontCommand{\sf}{\normalfont\sffamily}{\mathsf}
\DeclareOldFontCommand{\tt}{\normalfont\ttfamily}{\mathtt}
\DeclareOldFontCommand{\bf}{\normalfont\bfseries}{\mathbf}

\usepackage{enumitem}
\newlist{alphalist}{enumerate}{1}
\setlist[alphalist,1]{label=\textbf{(\alph*)}}
\usepackage{longtable}
\usepackage{lipsum} 
\usepackage{hyperref}
\usepackage{amsmath}

\hypersetup{
    colorlinks=true,
    linkcolor=green,
    citecolor=blue,
    urlcolor=blue,
}

\hypersetup{
        pdftitle={On the Statistical Interpretation of Discoveries in LHC Data},
        pdfsubject={On the Statistical Interpretation of Discoveries in LHC Data},
        pdfauthor={S. V. Chekanov and E. Wiek},
        pdfkeywords={Standard Model constants, quarks, vector bosons, genetic programming, artificial intelligence}
    }

\newcommand{\zgl}{$Z_{\,\mathrm{GL}}$ }
\newcommand{\zbh}{$Z_{\,\mathrm{BH}}$ }
\newcommand{\zlc}{$Z_{\,\mathrm{LC}}$ }

\journal{HEP-ANL-203752}
\begin{document}

\begin{frontmatter}


\title{On the Statistical Interpretation of Discoveries in LHC Data}

\author[a]{S.~V.~Chekanov}
\affiliation[a]{organization={HEP Division, Argonne National Laboratory},
            addressline={9700 S.Cass},
            city={Lemont},
            postcode={60516},
            state={IL},
            country={USA}}
\ead{chekanov@anl.gov}

\author[b]{E.~J.~Weik}
\affiliation[b]{organization={Department of Physics, New York University},
            addressline={726 Broadway},
            city={New York},
            postcode={10003},
            state={NY},
            country={USA}}
\ead{eweik@nyu.edu}

\tnotetext[t1]{Preprint HEP-ANL-203752, May 16, 2026}

\begin{abstract}
We examine discovery criteria at the Large Hadron Collider (LHC) within a model-independent framework, with particular emphasis on the statistical signatures of new physics. This study is motivated by the recent shift from model-specific searches based on a small number of distributions to broad, model-agnostic strategies, which offer substantially greater sensitivity to unexpected phenomena.
We revisit the well-known criterion of a local statistical significance of $5\,\sigma$ for the observation of new phenomena in invariant-mass distributions and discuss how this threshold should be  modified to account for look-elsewhere effects arising not only from multiple bins within a given distribution, but also from the simultaneous consideration of multiple distributions. We present a simple but statistically conservative relation between local and global significances in the presence of multiple invariant-mass distributions at the LHC, which can serve as a useful first approximation for planning future measurements.
\end{abstract}

\begin{keyword}
Standard Model, LHC, statistics

\PACS{14.80.-j, 13.85.Rm, 29.85.Fj, 12.38.Qk, 02.70.Rr}

\end{keyword}

\end{frontmatter}




\section{Introduction}
\label{sec:intro}

Since the first collision data were collected at the LHC in 2009, no statistically significant deviation from Standard Model (SM) predictions has been observed. 
This motivates the use of broad, model-agnostic search strategies that probe many distributions (see a recent review on anomaly detection \cite{Belis:2023mqs}), rather than focusing on a limited set of signal distributions associated with specific scenarios of physics beyond the SM.
This conceptual shift raises important statistical questions about how to claim a discovery when multiple distributions are examined simultaneously.

To date, no deviation from the SM with a significance greater than $5\,\sigma$ has been observed at the LHC, which is the conventional local-significance threshold for claiming the observation of new phenomena. In this paper, we use $Z$ to denote the significance, expressed as the number of standard deviations $\sigma$. For reference, a $5\, \sigma$ significance corresponds to a probability of about 1 in 3.5 million that the observed signal arises from a fluctuation of the known background.

In this work, we focus on resonance-like signals, namely localized excesses above a smooth background. Invariant-mass distributions $m_{12} = \sqrt{(p_1 + p_2)^2}$, where $p_1$ and $p_2$ are the four-momenta of particles (or jets) 1 and 2, have long been regarded as a golden channel for discovering new physics. The $J/\psi$ discovery (1974), the $\Upsilon$ discovery (1977), the Higgs discovery, and vector-boson studies serve as good examples. The search for ``bumps'' on smoothly falling with $m_{12}$ distributions is well established and requires relatively limited reliance on Monte Carlo simulations for modeling the SM processes. In many cases, the SM background can be described by monotonically falling functions above the Sudakov peaks.

The study of multiple invariant-mass distributions raises the question of whether, given the large number of studies published by the LHC experiments, one should expect to observe such excesses somewhere in the data purely by chance.
This effect, known as the ``look-elsewhere effect'', should account not only for the probability of observing excesses anywhere within the mass range \cite{Gross:2010qma}, but also for the number of distributions examined.

This issue is generally not addressed in current LHC studies of invariant-mass distributions, even in publications that analyze multiple distributions within a single study and conduct model-agnostic searches, i.e., without relying on specific theoretical predictions (see the discussion below).
In a broader historical context, accounting for the look-elsewhere effect is even more challenging, because it is difficult to determine how many invariant-mass distributions have been examined across the full LHC literature and what correlations may exist among the published results.

In this work, we address this question using a simplified, statistically conservative approach based on toy pseudoexperiments.


\section{Results}
\label{search}

In searches for new heavy resonances, both the ATLAS and CMS collaborations typically adopt the following strategy: First, the invariant mass $m_{12}$ of two objects, say two jets, is reconstructed. Then, a background hypothesis must be constructed, usually in terms of some analytic function, such as: 
$f(x) = p_1 (1 - x)^{p_2} x^{p_3 + p_4\ln x + p_5 \ln^2 x}$,
where $x \equiv m_{12} /\sqrt{s}$, $\sqrt{s}$ is the centre-of-mass energy of the $pp$ collisions,  and $p_i$\ ($i=1,\cdots 5$) are five free parameters. This function decreases monotonically with increasing $x$, reflecting the combined effect of parton distribution functions (PDFs), QCD partonic matrix elements, and the overall phase space.
To determine whether the data deviate from the
background-only hypothesis, the BumpHunter (BH) algorithm \cite{bumphunter} is most commonly used. This test calculates the ``local'' significance (\zlc) of any excess found 
in intervals scanned across the binned invariant-mass distribution.
It is derived from a local $p$-value, calculated from a hypothesis-test statistic, using the one-sided Gaussian conversion $Z = \Phi^{-1}(1 - p)$, where $\Phi$ is the cumulative distribution function of the standard normal distribution. 
Then, a look-elsewhere effect~\cite{Gross:2010qma} is taken into account 
by combining the separate hypothesis tests into a single test, and calculating the minimum $p$-value among all tests.
A global $p$-value is then calculated and transformed into statistical significance assuming that bin-by-bin
fluctuations of the data follow a Poisson distribution.
Pseudo-experiments are then used to determine the most significant local excess and, finally, a global significance $Z$ is calculated using the one-sided Gaussian conversion. 
We denote this significance by ``BH'', $Z_{\mathrm{BH}}$. The \zbh value provides the appropriate metric for the significance of the observation when accounting for the look-elsewhere effect arising from multiple bins, all possible bump locations, and varying widths. In the case of invariant masses, \zbh=$3\,\sigma$ is typically interpreted as evidence, while a significance of $5\,\sigma$ is conventionally considered a discovery.

As a specific example, let us consider an ATLAS publication \cite{ATLAS:2020zzb} where several invariant masses of two jets under different selections were studied. The largest deviation from the background-only hypothesis identified by the BH algorithm was near 1.3~TeV, with a local significance of $2.8\,\sigma$. Taking into account the look-elsewhere effect, the global $p$-value for the  largest deviation is $0.3$, leading to a significance of $0.5\,\sigma$ \cite{ATLAS:2020zzb}.
In our notation, \zlc= $2.8\,\sigma$ leads to  \zbh = $0.5\,\sigma$. Note that the latter value also incorporates systematic uncertainties associated with an alternative background shape. 

The above example demonstrates that the local significance of $3\,\sigma$ for a localized deviation from the background hypothesis translates into no meaningful evidence for a signal  assuming  the look-elsewhere effect for a single distribution.  Note that this reduction depends on the binning and the scan procedure. 

It is not unreasonable to expect that the true global significance (denoted by $Z_{\mathrm{GL}}$) is even smaller than $Z_{\mathrm{BH}}$, since  its value is further penalized by the number of studied distributions.
For example, in \cite{ATLAS:2020zzb}, this penalty reflects the 3 invariant masses. In \cite{ATLAS:2022ddh}, the penalty factor comes from 4 distributions, while 
it should reflect 9 distributions in  \cite{ATLAS:2023ixc}.
These publications do not quote global significances \zgl that reflect multiple distributions, but they typically provide \zbh values.

Naively, as a first approximation, one can estimate \zgl just by merging all examined histograms into a single histogram, processing them with the BH, and calculating \zbh which should approximate the true global \zgl value (here we again assume that all histograms under study are not strongly correlated). 
This simplistic method is not possible, because finding a discontinuous background function for merged histograms covering different scales is difficult.

\begin{figure*}[ht] 
         \centering
        \includegraphics[width=0.9\textwidth]{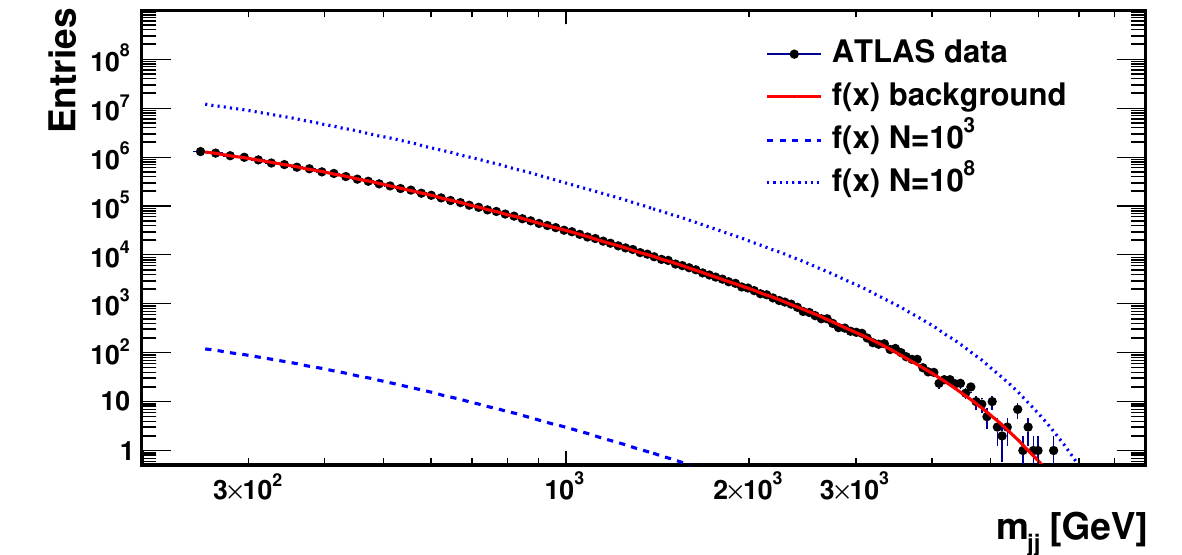}
    \caption{
An illustration of the two-jet invariant mass ($m_{jj}$) from the ATLAS study \cite{ATLAS:2020zzb}, compared with the function $f(x)$ (see text). Also shown are the two extreme normalization cases used in this study, $N=10^3$ and $N=10^8$.
 }
    \label{fig:0}
\end{figure*}

To illustrate the possible size of this additional trial factor, we will estimate \zgl in a much more realistic case using toy pseudo-experiments.
For practical purposes, we assume  that all two-particle invariant masses are studied for well-identified particles (or jets), since this is typically the first analysis one performs in such experiments. Let us consider only a few objects, such as jets, $b$-tagged jets, electrons, muons, and photons. The total number of unique two-particle combinations that can be formed from these objects, allowing for repeated object types such as two-jet invariant masses ($m_{jj}$), and treating leptons with different charges as separate particles, is 28.

There is little doubt that the ATLAS and CMS experiments have studied far more than 28 invariant masses. For example\footnote{We do not give the full list of ATLAS and CMS citations because of page limits.}, they have considered $Z/W$ bosons \cite{CMS:2014mws}, $\tau$-leptons \cite{CMS:2016xbv}, different trigger conditions \cite{ATLAS:2025okg} or dijets in events  with other particles/jets  \cite{ATLAS:2024qqm}, as well as three- and four-body invariant masses \cite{ATLAS:2022ddh}. 
Some publications reported studies of hundreds of different event categories  \cite{ATLAS:2018zdn,CMS:2020zjg}.
Generally, the expectation for two-body invariant masses selected in events within different exclusive event classes at the LHC is larger than 50,000 \cite{Chekanov:2023dby}.
In this paper, we first consider 28 invariant-mass distributions and then examine how \zgl changes as the number of distributions increases.

The natural question then arises: what happens if an analyst examines 28 independent histograms? What would be the global statistical significance, $Z_{\mathrm{GL}}$, of an observed ``signal'' arising purely from statistical fluctuations when accounting not only for multiple bins within each histogram, processed by the BH algorithm, but also for the fact that 28 independent histograms are inspected?

Let us provide a simple estimate using toy pseudo-experiments.
For each pseudo-experiment, the template function $f(x)$, discussed above, was used to create 28 random distributions by fluctuating each bin according to Poisson statistics, $N^{\mathrm{toy}} \sim \mathrm{Poisson}\!\left(N^{\mathrm{bkg}}_{b}\right)$, where  $N^{\mathrm{bkg}}$ is created from $f(x)$. It was assumed that the draws were mutually independent across bins. 
The bin widths were chosen to be consistent with the dijet-mass resolution, as in  \cite{ATLAS:2020zzb,ATLAS:2022ddh,ATLAS:2023ixc}.
The fit parameters used in $f(x)$ were taken  from the description of dijet invariant masses \cite{ATLAS:2020zzb}, which is representative of a typical dijet-mass distribution in $pp$ collisions, strongly shaped by the PDFs.
The parameters were $p_2=17.94$, $p_3=4.54$, $p_4=2.26$, $p_5=0.26$ for 
$\sqrt{s}=13$~TeV, which should broadly represent a typical monotonically falling distribution. The normalization parameter $p_1$ was set to different values to obtain the number of entries $N$ in the distributions, from $10^3$ to $10^8$, which will cover most of the LHC invariant-mass distributions. The minimum range was fixed at 0.25 TeV. Figure~\ref{fig:0} illustrates the ATLAS data  \cite{ATLAS:2020zzb}, the $f(x)$ function with the parameters mentioned above, and the two extreme cases of the normalization used for our studies. The maximum range for counting statistics depends on normalization and varies from 1.6 TeV (for $N=10^3$ entries) to about 7.2 TeV for $N=10^8$.  In total, 100 million toy pseudo-experiments were created for all $N$. 

For each choice of normalization, the BH tool was used to estimate how often any of the 28 distributions produced a local enhancement with a significance greater than a given \zlc at any position above 250~GeV under the null background hypothesis. 
The calculations were performed in a step of $0.5\,\sigma$, from \zlc$=3\,\sigma$ to \zlc$=6\,\sigma$.
The BH algorithm requires the excess to span more than a single histogram bin. This commonly used criterion is sensitive to a broad range of possible resonant phenomena, from narrow states with widths comparable to the dijet-mass detector resolution to broad states.
Then we counted these rare events and divided by the total number of pseudo-experiments. This ratio represents the ``global'' $p$-value, i.e., the probability of observing a localized signal  with a strength \zlc $\geq 5\,\sigma$, as reported by the BH tool. Then this $p$-value was transformed into the \zgl value as described above. We disabled the calculation of \zbh  (for a single histogram) within the BH procedure, which is not required in this approach. For simplicity, we treat all histograms as independent. This serves as a first approximation before considering correlations between the invariant masses, which are beyond the scope of this study.

The background yields are expected to be different for different invariant masses constructed from jets, leptons and photons. Therefore, we repeated the toy pseudo-experiments after varying the normalization parameter ($p_1$) of $f(x)$ to represent different normalizations (denoted with the letter $N$) of two-body invariant masses, from $N=10^3$ to $N=10^8$.
In practice, the 28 histograms are expected to have different normalizations depending on the object type, but such variations should be well covered by our choice to vary the normalization while applying the same normalization shift to all histograms. The numbers of invariant masses used in the tests are denoted by the letter $D$. All distributions were assumed to have the same shape. It was determined that the shape-changing effect has a much smaller impact on the results than the effect on $N$ (see the discussion below).

\begin{figure}[h!] 
         \centering
        \includegraphics[width=\linewidth]{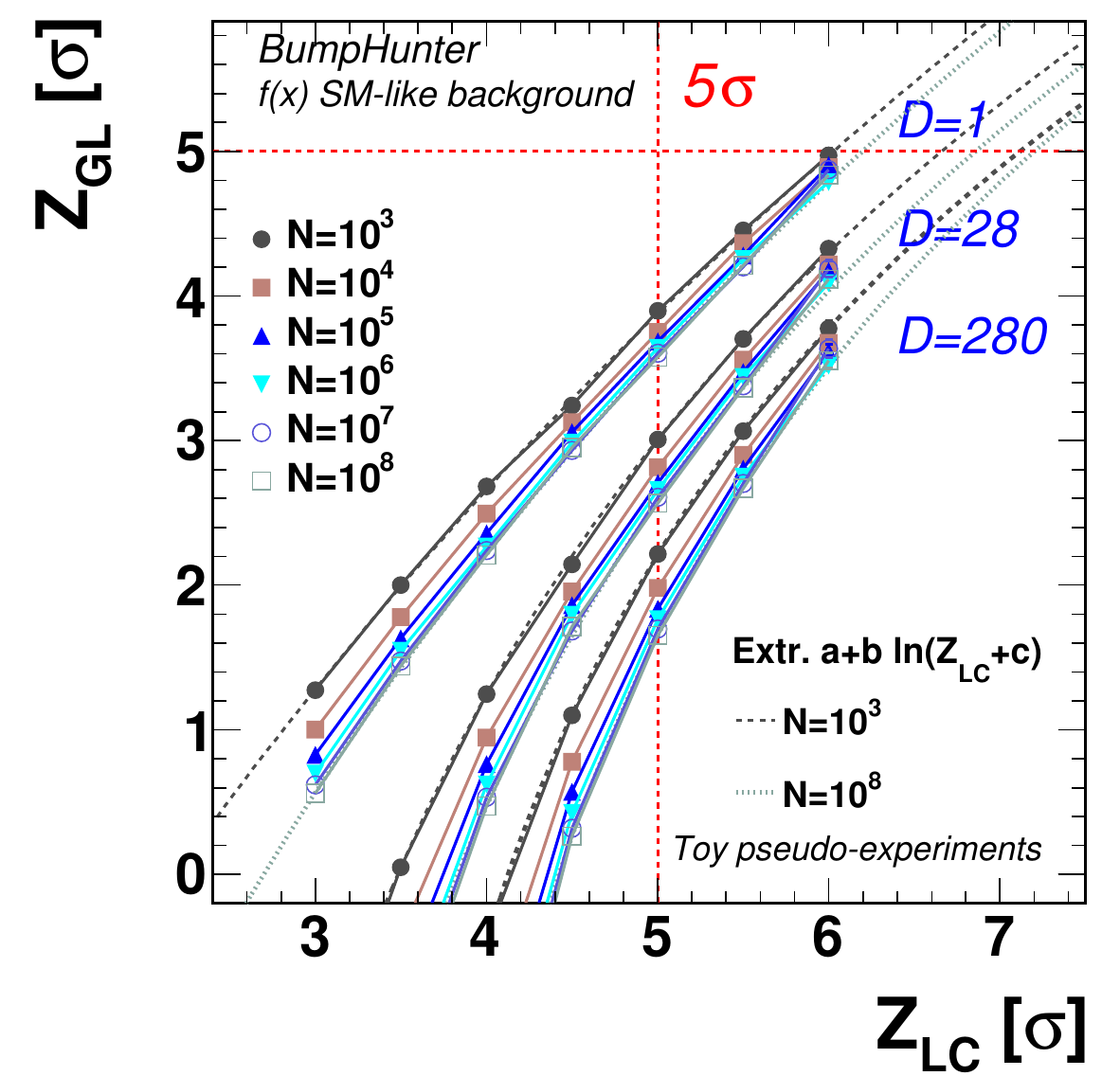}
    \caption{
Relationship between \zlc and \zgl for different normalizations and different numbers of invariant-mass distributions (denoted by $D$), namely 1, 28, and 280. The normalization of the background template, $f(x)$, used in the pseudo-experiments ranges from $N=10^3$ to $N=10^8$, with the results shown separately for each normalization. The BH tool was used to identify localized enhancements while accounting for look-elsewhere effects. Vertical error bars indicate the statistical uncertainties from the toy pseudo-experiments. The calculations are performed  with a step of $0.5\,\sigma$ from \zlc$=3\,\sigma$ to \zlc$=6\,\sigma$. The values of \zgl below 0 are not shown. The solid lines are used to guide the eye.  
The black dashed and dotted lines show the $\chi^2$ fits using the function $a+b \ln (Z_{\mathrm{LC}}+c)$, employed to extrapolate the data to large values of \zlc for the two extreme cases, $N=10^3$ and $N=10^8$.}
    \label{fig:1}
\end{figure}

Figure~\ref{fig:1} shows the dependence of \zgl on \zlc for $D=28$ invariant-mass distributions  for different normalizations from $N=10^3$ to  $N=10^8$.
According to the calculation, an observation of a \zlc$\geq 5\,\sigma$ signal in any of the 28 histograms, at any position and width, corresponds to a global significance of only \zgl$\simeq 2.5 - 2.8\,\sigma$, with a small variation depending on the number of entries discussed above. 

The global significance \zgl decreases as $N$ increases, because the maximum invariant mass range grows with $N$, reaching approximately 7~TeV for $N = 10^8$. This expansion enhances the bin-related look-elsewhere effect.

For all values of $N$ at \zlc = $5\,\sigma$, the observed \zgl by the BH algorithm does not meet the conventional criterion of $3\,\sigma$ for evidence for new resonant phenomena when the look-elsewhere effect is taken into account.

It should be noted that a local excess identified by the BH algorithm can arise more easily from statistical fluctuations than a single-bin excess. This is because BH scans many possible intervals and can combine several moderate upward fluctuations in neighboring bins, as expected for a signal-like structure. Therefore, observing a local significance of  \zlc=$5\,\sigma$ at some position in a distribution is generally more likely with BH than in a single bin of a histogram with independent bins.

Figure~\ref{fig:1} also shows the dependence of \zgl on \zlc for $D=1$ (i.e., a single distribution). In this case, the look-elsewhere effect is bin-related , reflecting the unknown position (and width) of a fluctuation in a single histogram during the BH scan. As a result, \zgl is closer to \zlc than for $D=28$.
Increasing the number of histograms by a factor of 10, i.e., to $D=280$ invariant masses, reduces the global significance, highlighting the severity of the look-elsewhere effect. 
In this case,  \zlc$=5\,\sigma$ leads to \zgl$\simeq 1.5\,\sigma$.

The same Fig.~\ref{fig:1} also shows the extrapolation results for the two extreme cases, $N=10^3$ and $N=10^8$, extended to larger values of $Z_{\mathrm{LC}}$. These extrapolations were obtained by fitting the toy pseudo-experiment data with the function $a+b \ln (Z_{\mathrm{LC}}+c)$, where the parameters $a$, $b$, and $c$ were determined from a $\chi^2$ minimization of the simulated data. The fit quality was acceptable, with $\chi^2/\mathrm{ndf} < 1.5$. 
Another function, a saturating exponential of the form $a - b \exp(-c Z_{\mathrm{LC}})$, provided a slightly poorer fit.

It was verified that the results do not depend strongly on the details of the shape of the background function $f(x)$, as long as it follows the typical ``PDFs-like'' monotonic decrease and the maximum range does not exceed 7~TeV. As a cross-check, the invariant-mass shape was taken from one of the distributions shown in \cite{ATLAS:2023ixc}, which exhibits a less steep monotonic fall due to the applied anomaly-detection selection. 
It was concluded that the variation arising from changing $N$ from $10^3$ to $10^8$ fully dominates the spread shown in Fig.~\ref{fig:1}, and therefore is not expected to depend significantly on the shape. This study does not account for other sources of systematic variation that are typically considered in experimental papers.

It should be emphasized that this paper concerns the largest deviation observed in at least one invariant-mass distribution under the null hypothesis,  i.e. it  is defined with respect to statistical fluctuations of SM-like backgrounds only. Correlations among invariant-mass observables, arising for example from event kinematics, can produce weaker ``reflections'' of the leading fluctuation in other distributions. However, because the analysis is based on the maximum local excess found in the signal scan, such secondary correlated deviations do not change the definition of the test statistic used here. The interpretation of a significant excess in terms of conditional probabilities for correlated signatures in other channels is a separate, model-dependent step relevant to physics interpretation after a candidate signal has been identified, rather than to the null-hypothesis look-elsewhere estimate itself.

It should also be noted that this work focuses on the statistical treatment of localized excesses identified in model-agnostic scans of invariant-mass distributions, where the excess is bounded on both sides and can be characterized as a resonance-like feature. Other classes of anomalies, including broader distortions in mass tails or correlated deviations across multiple mass distributions, may be more naturally interpreted within effective-field-theory frameworks \cite{Brivio:2017vri}. However, such cases require a separate discussion and lie beyond the scope of this paper.

So far, no \zlc$\geq 4.5\,\sigma$ signal has been observed in the CMS or ATLAS data \cite{lhcex}. However, for purely statistical reasons, the observation of a local fluctuation with this significance should not be regarded as extremely surprising: Even for a relatively small number of invariant-mass distributions (28), a localized enhancement with \zlc$\simeq 4.5\,\sigma$, as seen by the BH, is not unexpected, since such enhancements should lead to \zgl$\simeq 1\,\sigma$. 
It should be even more likely to observe statistical fluctuations at the $4.5\,\sigma$ level when considering more than 28 quasi-independent distributions studied by the LHC experiments under different event selections to optimize sensitivity to various models beyond the SM. 
If one retrospectively examines previous analyses, the absence of published excesses of this strength at the LHC suggests that the public results may not follow the usual statistical pattern. A possible explanation of this effect is the inclusion of instrumental uncertainties, which can distort fluctuations due to purely statistical reasons, positive correlations between distributions, which may effectively reduce the number of independent trials,  and the less traditional use of blinding techniques \cite{Chekanov:2021kwl}, along with other effects.

Another question may arise: what would be the local significance of localized enhancements in invariant-mass distributions to constitute a compelling case for new physics in the future? The answer depends on a careful assessment of the look-elsewhere effect, accounting for the many studies conducted across different mass ranges.  As discussed above, exploration of many studied distributions and their correlations represents a significant challenge, plus computer simulations for the pseudo-experiments  beyond $6\,\sigma$ localized enhancements are very computationally demanding.

The study presented in this paper is only illustrative and should not be interpreted as a universal conversion rule between local and global significance. The results depend on the number of distributions studied, histogram correlations, background modeling, and other factors. For example, histograms may be positively correlated. Such correlations are generally expected, as an increase in event multiplicity typically leads to more entries at all invariant masses across different object types.
Due to these correlations, the effective number of independent trials can be smaller than the raw counts (28 or 280) considered in this study. If such positive correlations are taken into account, the statistical expectations for global significance will be less conservative than those presented here (i.e. \zgl will be larger for fixed $N$ and $D>1$). 
Nevertheless, to a first approximation, Fig.~\ref{fig:1} already serves as a useful conservative ``look-up'' reference for assessing the potential significance of observations during experimental preparation and for estimating the corresponding global significance for ``wide'' searches involving many distributions.

On a practical note, a simple way to apply the results of this paper is to estimate the effective number of distributions and their normalization, and then use Fig.~\ref{fig:1} to obtain an approximate global significance. If Fig.~\ref{fig:1} is not convenient, for example because it is defined only for coarse choices of the number of distributions (such as 28 or 280), one may instead use the numeric program \cite{GlobalMultiMassGitHub}, which was used in this study and allows arbitrary choices of both the normalization and the number of distributions.

\section{Conclusions}

The studies presented in this paper  highlight the statistical complexity of LHC data analysis when it comes to claiming a discovery. Our central estimate is based on 28 distributions, 
derived from the number of two-particle invariant-mass combinations that can be constructed to cover the most prominent invariant masses in inclusive collision events. For the LHC, we consider this a conservative estimate, since the experiments have examined an even larger set of distributions under varied selection criteria, including those extending beyond two-particle spectra. 
Moreover, treating these distributions as independent is expected to make the calculation of $Z_{\mathrm{GL}}$ conservative as well, which is an appropriate approach when claiming major new phenomena.
Under these conditions, the conventional local $5\,\sigma$ requirement leads to \zgl$<3\,\sigma$.
Thus, this criterion no longer provides sufficient evidence for new phenomena in a model-agnostic exploration of invariant masses. The relevance of this rule is further reduced as the number of tested distributions increases. In such situations, the more stringent \zgl$>5\,\sigma$ requirement is more appropriate, which corresponds to a substantially larger \zlc significance. Extrapolating the relations shown in Fig.~\ref{fig:1} suggests that \zlc should exceed $7\,\sigma$ when more than 28 distributions are considered.

It should also be noted that the latter threshold can revert to the usual \zlc=$5\,\sigma$ if the observation is guided by a theoretical prediction for a specific invariant mass and well-defined signal characteristics, thereby effectively removing the look-elsewhere effect.

\vspace{0.5cm}

{\it Data Availability}. The code used in this paper is available on GitHub \cite{GlobalMultiMassGitHub}.

{\it Acknowledgments:} 
In memory of Dr. James D.~Bjorken (1934 - 2024), who, during a private discussion with one of the authors (S.C.) at the ISMD13 conference, raised concerns about statistical features of published LHC results. 
The submitted manuscript has been created by UChicago Argonne, LLC, Operator of Argonne National Laboratory (“Argonne”). Argonne, a U.S. Department of Energy Office of Science laboratory, is operated under Contract No. DE-AC02-06CH11357.  Argonne National Laboratory’s work was 
funded by the U.S. Department of Energy, Office of High Energy Physics under contract DE-AC02-06CH11357.

\bibliographystyle{elsarticle-num}
\bibliography{references}





\end{document}